\documentclass[aps,prc,twocolumn,groupedaddress,showpacs]{revtex4}

\usepackage{graphicx}

\newcommand{\dt}{\partial t}
\newcommand{\Dt}{\mbox{d} t}

\newcommand{\der}{\partial}

\newcommand{\ddt}{\frac{\der}{\dt}}

\newcommand{\Fop}{\hat{F}}

\newcommand{\bra}[1]{\langle #1 \vert}
\newcommand{\ket}[1]{\vert #1 \rangle}
\newcommand{\bran}[1]{\langle #1}
\newcommand{\com}[1]{\left[ #1 \right]}

\newcommand{\crea}{\hat{a}^\dagger}
\newcommand{\anni}{\hat{a}}

\newcommand{\hbc}{\hat{\beta}^\dagger}
\newcommand{\hba}{\hat{\beta}}

\newcommand{\jmat}{\mathbf{j}}
\newcommand{\Jso}{\mathbf{J}}

\newcommand{\Jsob}{\mathbf{J}}

\newcommand{\ofR}{(\mathbf{r})}
\newcommand{\ofr}{(r)}
\newcommand{\oft}{(t)}

\newcommand{\Smat}[4]{\left( \begin{array}{cc} #1 & #2 \\ #3 & #4 \end{array} \right)}
\newcommand{\Svec}[2]{\left( \begin{array}{c} #1 \\ #2 \end{array} \right)}
\newcommand{\Hbogo}{\mathcal{H}}
\newcommand{\Rbogo}{\mathcal{R}}

\newcommand{\Ox}[1]{^{#1}\mbox{O}}
\newcommand{\Ca}[1]{^{#1}\mbox{Ca}}

\begin{document}


\title{Pairing Vibrations Study with the Time-Dependent Hartree-Fock-Bogoliubov theory}

\author{B. Avez} 
\email{benoit.avez@cea.fr}
\affiliation{CEA, Irfu, Service de Physique Nucl\'eaire,\\ 
Centre de Saclay, F-91191 Gif-sur-Yvette, France.}
\author{C. Simenel} 
\email{cedric.simenel@cea.fr}
\affiliation{CEA, Irfu, Service de Physique Nucl\'eaire,\\ 
Centre de Saclay, F-91191 Gif-sur-Yvette, France.}
\author{Ph. Chomaz}
\affiliation{GANIL (DSM-CEA/IN2P3-CNRS), B.P. 55027, 
F-14076 Caen cedex 5, France}
\affiliation{CEA, Irfu/Dir, Centre de Saclay, F-91191 Gif-sur-Yvette, France.}
\date{\today}

\begin{abstract}
We study pairing vibrations in $^{18,20,22}$O and $^{42,44,46}$Ca nuclei
solving the time-dependent Hartree-Fock-Bogoliubov equation 
in coordinate space with spherical symmetry.  
We use the SLy4 Skyrme functional in the normal part of the 
energy density functional and a local density dependent 
functional in its pairing part.
Pairing vibrations are excited by two-neutron transfer operators.
Strength distributions are obtained using the Fourier transform of the time-dependent
response of two-neutron pair-transfer observables in the linear regime.
Results are in overall agreement with quasiparticle random phase approximation 
 calculations for Oxygen isotopes, though differences appear when 
 increasing the neutron number.
 Both low lying pairing modes and giant pairing vibrations (GPV) are discussed.
The GPV is observed in the Oxygen but not in the Calcium isotopes.
\end{abstract}


\maketitle

\section{Introduction}
Energy density functional (EDF) approaches like the Skyrme-Hartree-Fock 
model~\cite{sto07,BenderHeenenReinhard} have proved to be successful to 
describe bulk properties of nuclei over the nuclear chart~
\cite{BrinkVautherin}. Recent computer improvements allow large scale 
EDF calculations of nuclear structure~\cite{ben05a,ben05b} and 
reactions~\cite{UmarOberacker, CSBA}. 
In these approaches, one restricts the many-body 
wave-functions to a subset of the Hilbert space 
 on the one hand and guess a nuclear EDF on the other hand.
A commonly used technique is to break symmetries to enrich 
the variational sub-space and improve the description of nuclear structure.
As an example, breaking gauge invariance associated 
to particle number conservation yields the Hartree-Fock-Bogoliubov 
(HFB) formalism~\cite{Hartree, Fock, Bogoliubov}.
This technique allows for the description of superfluidity in 
ground states of open-shell nuclei. 

Extensions to treat collective excitations in the presence
of pairing correlations are possible in the framework
of the Quasiparticle Random Phase Approximation (QRPA)
which has been widely used in nuclear structure 
studies~\cite{eng99,ben02,cQRPA,fra05,ter05,ter06,per08}. 
This approach and its zero pairing counterpart (RPA) 
give reasonable estimates  of giant resonances
though improvements are necessary to reproduce fine structures~\cite{lac04}. 
In fact, the QRPA can be obtained from the linearization of the 
time-dependent Hartree-Fock-Bogoliubov (TDHFB) equation
which provides a self-consistent evolution of an independent quasiparticles state.
There is a consistency requirement that QRPA and the static limit of TDHFB
should use the same effective interaction.
This is a natural feature of TDHFB that  the same EDF
can be easily used in the static and dynamical calculations
thanks to the structure of the TDHFB equation. 
This is not always the case in (Q)RPA calculations where 
spin-orbit and Coulomb parts of the
residual interaction are often omitted 
(see discussion in~\cite{nak05} and references therein),
which may affect collective modes~\cite{ter05,colo04,per05,sil06}.

Unlike in condensed matter where, for instance, TDHFB has been applied
 to study dynamics of Bose-Einstein condensates~\cite{Holland, bul05},
explicit time evolutions of nuclei including pairing are sparse
and usually limited to the BCS ansatz of superfluidity with simple functionals
~\cite{neg78, cus79, cus80} or to simple systems~\cite{BlockiFlocard}.
Only recently, a numerical method of solving TDHFB with the Gogny interaction
~\cite{dec80} has been proposed to study quadrupole oscillations using a 
harmonic oscillator basis~\cite{HashimotoNodeki}.

At the limit where pairing is neglected, however,
extensive calculations of nuclear dynamics have been performed
 using the time-dependent Hartree-Fock (TDHF) formalism introduced 
 by Dirac in 1930~\cite{dirac}. In this approach, one considers 
 the dynamics of independent particles in a self-consistent mean-field generated 
 by all the others.
 The use of Skyrme EDF~\cite{sky56} allowed recent realistic 
 calculations of both collision 
mechanisms~\cite{bon76, Negele, sim01, sim04, UmarOberacker, mar06, sim07, CSBA} 
and giant resonances~\cite{sim03,nak05,uma05,mar05}.
For instance, TDHF has been used to study anharmonicities in collective motion 
which are beyond the RPA range of applications~\cite{sim03, rei07}. 
It is then an appealing challenge to repeat these works using TDHFB 
to investigate the role of pairing correlations in nuclear dynamics.
In particular, ''How do they affect low-energy reaction 
mechanisms~?'' is still an open question.
However, full three-dimensional TDHFB codes for collisions are still prohibitive at moment. 
For nuclear structure purposes, pairing vibrations 
and rotations~\cite{BohrMottelson, RnS, BesBroglia}
also demand theoretical investigations to reach realistic predictions~\cite{das06}.
Of particular interest are high-lying pairing collective modes 
like giant pairing vibrations (GPV)~\cite{BrogliaHighLying}. 
These modes can be viewed as coherent sums 
of two-quasiparticle excitations across a major shell.
They are probed by two-nucleon transfer 
reactions~\cite{RipkaPadjen,BrogliaHansenRiedel}. 
The GPVs are still unobserved experimentally but they are 
experiencing a renewed interest
since recent theoretical developments predicted that the use of 
radioactive ion beams could provide better conditions 
for their studies~\cite{Fortunato, VonOertzenVitturi, Khan}.

In this article, we solve for the first time the TDHFB equation with a 
full Skyrme functional in the normal part of the EDF
and a local density dependent one in its pairing part. 
As a first application, we investigate pairing vibrations 
 using the linear response theory. 
Our model is then equivalent to a fully consistent QRPA including
spin-orbit and Coulomb interactions also in the particle-hole channel. 
In section~\ref{sec:formalism}, 
we recall the TDHFB formalism and the choice of the EDF. 
We present in section~\ref{sec:numerical}  
numerical implementations in spherical symmetry.
Then, we discuss in section~\ref{sec:linear} our choice of 
observables associated to pairing vibrations in the framework
of the linear response theory.
Finally, we present the results for  $\Ox{18,20,22}$ and $\Ca{42,44,46}$
 isotopes in section~\ref{sec:results}
before to conclude in section~\ref{sec:conclusion}.

\section{Formalism \label{sec:formalism}}

\subsection{TDHFB equation}

The TDHFB equation can be derived starting from the action
between an initial and final time $t_i$ and $t_f$
\begin{equation}
S = \int_{t_i}^{t_f} \Dt \, \, \bra{\Psi\oft} i \hbar \ddt -\hat{H} \ket{\Psi\oft}
\label{PVfunc}
\end{equation}
and writing the variational principle $\delta S=0$ 
in the sub-space of quasiparticle vacua.
For each state $\ket{\Psi}$ 
of this sub-space, one can find a basis of quasiparticle annihilators
$\{\hat{\beta}\}$ such that
 $ \hba_\mu \ket{\Psi} = 0$ for all $\mu$~\cite{RnS}.
The latter can be related to the particle creation and annihilation operators 
$\{\hat{a}^\dagger,\hat{a}\}$ through the Bogoliubov transformation~\cite{Bogoliubov}
\begin{equation}
\hat{\beta}_\mu = \sum_\nu \left( U^*_{\nu \mu} \hat{a}_\nu 
+ V^*_{\nu \mu} \hat{a}^\dagger_\nu\right)
\label{eq:betadag}
\end{equation}
where the matrices $U$ and $V$ are such that the quasiparticle operators 
$\{\hat{\beta}^\dagger,\hat{\beta}\}$
fulfill the canonical anti-commutation rules for fermions.

The variational principle leads to the TDHFB equation~\cite{BlaizotRipka}
\begin{eqnarray}
i \hbar \ddt \Rbogo= \com{\Hbogo, \Rbogo}.
\label{eq:TDHFB}
\end{eqnarray}
The generalized one-body density matrix reads
\begin{eqnarray}
\Rbogo\oft= \Smat{ \rho\oft }{ \kappa\oft }{ -\kappa^*\oft }{ 1-\rho^*\oft },
\end{eqnarray}
where $\rho_{\mu\nu}=\left<\psi| \crea_\nu \anni_\mu|\psi\right>$ 
are the matrix elements of the normal density
and  $\kappa_{\mu\nu}= \left< \psi|\anni_\nu \anni_\mu |\psi\right>$ 
are the elements of the pairing tensor.
The generalized one-body Hamiltonian $\Hbogo$ has a block structure
and can be writen in terms of the Hartree-Fock (HF) Hamiltonian $h$
 and the pairing field $\Delta$
 \begin{eqnarray}
\Hbogo=\Smat{ h }{ \Delta }{ -\Delta^* }{ -h^* },
\end{eqnarray}
where 
\begin{eqnarray}
h_{\mu\nu}=\frac{\delta \mathcal{E}[\rho,\kappa,\kappa^*]}{\delta \rho_{\nu\mu}} \mbox{~~and~~} 
\Delta_{\mu\nu}=\frac{\delta \mathcal{E}[\rho,\kappa,\kappa^*]}{\delta
 \kappa^*_{\mu\nu}} .
 \label{eq:field}
\end{eqnarray}
In the above HFB formalism, the functional $\mathcal{E}[\rho,\kappa,\kappa^*]$
is  the expectation value of the exact Hamiltonian $\hat{H}$
on the quasiparticle vacuum $\ket{\Psi}$. 

A more practical form of the TDHFB equation is found by
recasting these equations in terms of the quasiparticle components $U$ and $V$ 
introduced in Eq.~(\ref{eq:betadag})
\begin{eqnarray}
i \hbar \ddt \Svec{U_{\nu \mu}}{V_{\nu \mu}} = 
\sum_{\eta} \Smat{h_{\nu \eta}}{\Delta_{\nu \eta}}{-\Delta_{\nu \eta}^*}{-h_{\nu \eta}^*}
\Svec{ U_{\eta \mu} }{ V_{\eta \mu} }.
\label{eq:TDHFB_QPcomp}
\end{eqnarray}

\subsection{EDF approach}
\label{sec:EDF}

In nuclear physics, the above formalism should be modified, in particular
to take into account the short range repulsive part of the nuclear interaction
which makes the mean-field HFB approach irrelevant with the bare interaction.
This is a reason why one generally replaces $\mathcal{E}[\rho,\kappa,\kappa^*]$
by an effective EDF fitted on nuclear properties
without invoking directly the underlying exact Hamiltonian.
Moreover, in the spirit of the density functional theory~\cite{DFT_HK, 
DFT_KS, TDDFT_RG}, this procedure allows to include many-body correlations.

Let us decompose the total energy into kinetic, Skyrme, Coulomb and pairing parts
(see appendix~\ref{TDHFB_sphe} for an explicit expression of each component)
\begin{eqnarray}
\mathcal{E}=\mathcal{E}_{kin}+\mathcal{E}_{Sk}
+ \mathcal{E}_{Coul} + \mathcal{E}_{pair}.
\label{TDEDF}
\end{eqnarray}
We choose the SLy$4$ parameterization~\cite{Chabanat2} 
of the Skyrme functional~\cite{sky56} 
 including time-odd densities~\cite{Cranking}. 
The Coulomb energy includes direct and exchange terms. 
The latter is estimated using the Slater approximation~\cite{neg73}. 
The three first terms of Eq.~(\ref{TDEDF}) depend only on the normal densities.
 It is convenient to express the pairing energy $\mathcal{E}_{pair}$
using the anomalous density 
\begin{equation}
\tilde{\rho}_q(\mathbf{r}s,\mathbf{r}'s') = -2s'\kappa_q(\mathbf{r}s,\mathbf{r}'-s')
\label{eq:rhotilde}
\end{equation}
where $s$ the projection of the spin and $q$ the isospin.
We use a local pairing functional  
 (see, e.g.,~\cite{BenderHeenenReinhard} and references therein)
\begin{eqnarray}
\mathcal{E}_{pair} &=& \int \!\!\! \mbox{d}\mathbf{r}
\,\,\,  \frac{g}{4} \, 
\left[ 1 - \left(\frac{\rho_0\ofR}{\rho_c}\right)^\gamma \right]
 \sum_{q} \tilde{\rho}_q^*(\mathbf{r})\,  \tilde{\rho}_q(\mathbf{r}) 
\label{pairfunc}
\end{eqnarray}
where $\rho_q(\mathbf{r})=\sum_{s} \rho_q(\mathbf{r}s,\mathbf{r}s)$
and $\tilde{\rho}_q(\mathbf{r})=\sum_{s} \tilde{\rho}_q(\mathbf{r}s,\mathbf{r}s)$ 
are the local parts of the normal and anomalous densities with isospin $q$
respectively, $\rho_0(\mathbf{r})=\sum_{q} \rho_q(\mathbf{r})$ is the scalar-
isoscalar density and $g$ is the pairing coupling constant.
The parameters $\rho_c$ and $\gamma$ are adjusted to generate pairing
correlations preferably at the surface and/or in the bulk of the nucleus.
Such a pairing scheme yields pairs of nucleons of the same isospin
coupled to angular momentum zero. The simplicity of such an EDF made 
systematic three-dimensional HFB calculations over the whole nuclear 
chart possible~\cite{ben05a,ben05b}.

However, one has to face the divergence of local pairing densities~\cite{UVdiv}. 
It is then necessary either to 
regularize the equations by introducing a cutoff in the quasiparticle spectrum
or eventually to perform a more complex renormalization scheme 
(for an example based on the Thomas-Fermi approximation, see~\cite{bru99,Bulgac1}). 

In our calculations, we use a cutoff to regularize the TDHFB equation in a quasiparticle
energy window of $80$~MeV. 
This value allows two-quasiparticle excitations up to 160 MeV.
The pairing parameters are fitted to reproduce a neutron 
spectral gap of $1.25$~MeV in $^{120}$Sn, the pairing acting both in the bulk and at the 
surface of the nucleus~\cite{dob01}. The obtained pairing coupling constant is 
$g=-275.25$~MeV with the parameters $\rho_c = 0.32$~fm$^{-3}$ and~$\gamma=1$. 

\subsection{Particle number conservation \label{sec:particle}}

Due to the broken $U(1)$ gauge invariance associated to the 
particle number conservation, the HFB states are not eigenstate 
of the particle number operator $\hat{N}$. In static calculations, 
one adds a Lagrange multiplier~$\lambda$, 
interpreted as a chemical potential,
in order to  fix the number of particles in average.
In TDHFB dynamical simulations, however, the particle number obeys to
\begin{eqnarray}
  i \hbar \ddt \langle \hat{N} \rangle = 
    \mbox{Tr}\left( \kappa \Delta^* - \Delta \kappa^* \right).
  \label{eq:pairing_functional}
\end{eqnarray}
The definitions of $\Delta$  and  $\tilde{\rho}$ in Eqs.~(\ref{eq:field}) 
and~(\ref{eq:rhotilde}) respectively,
together with the choice of the pairing functional in Eq.~(\ref{pairfunc}) 
ensure that the right hand side of Eq.~(\ref{eq:pairing_functional}) vanishes.
As a consequence, we do not need to enforce the conservation of the average 
number of particles with a chemical potential in the TDHFB equation. 

Nevertheless, dropping this constraint in the dynamics 
induces a rotation of the Bogoliubov vacuum in gauge 
space~\cite{TDHFBBulgac}. 
In particular, the anomalous density of a stationary state will 
carry a phase $\exp{\left(-{2i\lambda t}/{\hbar}\right)}$.
Then, the ground state expectation values of observables which are linear 
in the anomalous density will evolve in time.
For instance, this is the case of the observable we use 
to study pairing vibrations (see  Eq.~(\ref{eq:Fdet})).
The time-Fourier analysis of such an observable in the linear response theory 
(see section \ref{sec:linear}) will then contain a spurious peak 
at an energy $\hbar \omega = 2 \lambda$.
This is a manifestation of the Goldstone mode due to 
the broken symmetry associated to particle number conservation.
In QRPA calculations, it induces a spurious mode at zero energy.
In order to avoid such a spurious mode in the linear response of TDHFB, we have to
 keep the static ground state chemical potential in the particle-hole field during 
the evolution.
Finally, we note that the chemical potential $\lambda$ is easy to compute  
only for nuclei with pairing. Thus we do not consider doubly magic nuclei
as initial states in the present work.

\section{Numerical implementation of TDHFB \label{sec:numerical}}

\subsection{Spherical symmetry}

The pairing functional defined in Eq.~(\ref{pairfunc}) couples only nucleons of the same isospin.
We can then focus on semi-magical nuclei for which 
the spherical assumption is a good approximation.
As a consequence, we solve the TDHFB equation using spherical symmetry.

Let us recast the problem with purely local fields in space, spin and isospin
using this symmetry. 
The total many-body wave-function being rotational invariant,
it is convenient to write the Bogoliubov transformation 
in the spherical basis using the standard notation for quantum numbers
$n$, $l$, $j$ and $m$ (we omit the isospin $q$ in the notation for simplicity)
\begin{eqnarray}
  \hbc_{nljm} = \sum_k(\mathcal{V}^{(ljm)}_{kn}(-1)^{j-m}\anni_{klj-m} + \mathcal{U}^{(ljm)}_{kn}\crea_{kljm}).
  \label{eq:beta_spher}
\end{eqnarray}
This definition 
ensures that the component $m$ of these quasiparticle operators 
transforms under rotation as  a tensor of rank $j$ (see Eq.~(8.79) 
of Ref.~\cite{BlaizotRipka}).

We choose to solve the TDHFB equation in coordinate space 
using quasiparticle wave functions 
$U_\nu( \mathbf{r} s)\equiv U_{ \mathbf{r}s, \nu}$ and
$V_\nu( \mathbf{r} s)\equiv V_{ \mathbf{r} s,\nu}$ 
 defined as components of quasiparticle spinors
\begin{eqnarray}
  \Svec{{U}_{nljm} \left( \mathbf{r} s\right)}{{V}_{nljm} \left( \mathbf{r} s \right)} = 
\Svec{ \bra{\mathbf{r} s}\hbc_{nljm}\ket{-} }{ \bra{-}\hbc_{nljm}\ket{\mathbf{r} s} },
  \label{eq:qp_spher}
\end{eqnarray}
where $\ket{-}$ is the particle vacuum.
The standard decomposition of single particle orbitals in spherical coordinates writes 
$\langle {\mathbf{r} s} \ket{nljm}=R_{nlj}(r) \,\Omega_{ljms}(\theta \phi)$.
The angular part is expressed in terms of Clebsch-Gordan coefficients 
and spherical harmonics 
$\Omega_{ljms} = \bran{l (m-s)\frac{1}{2}s}\ket{jm} Y_l^{m-s}$.
Defining the radial quasiparticle wave functions
$u_{njl}(r) = r \sum_k \mathcal{U}_{kn}^{(ljm)} R_{klj}(r)$ and
$v_{njl}(r) = (-1)^{l+1} r \sum_k \mathcal{V}_{kn}^{(ljm)} R^*_{klj}(r)$,
and using the property 
$\Omega^*_{lj-ms}= -2s (-1)^{m+l-j} \Omega_{ljm-s}$,
Eq.~(\ref{eq:qp_spher}) becomes
\begin{eqnarray}
 \Svec{{U}_{nljm} \left( \mathbf{r} s \right) }
      {{V}_{nljm} \left( \mathbf{r} s \right) } = 
 \frac{1}{r}\Svec{ u_{nlj}\left(r\right)\, \, \Omega_{ljms}\left(\theta \phi\right) }
                 { 2\sigma \, \, v_{nlj}\left(r\right) \, \, \Omega_{ljm-s}\left(\theta \phi\right) }. \nonumber
\end{eqnarray}

Following the same way as Dobaczewski {\it et al.} 
for the static HFB problem~\cite{NPA422}, 
we introduce  the anomalous field $\tilde{h}_q(\mathbf{r}s,\mathbf{r}'s')
= \frac{\delta \mathcal{E}[\rho,\tilde{\rho},\tilde{\rho}^*]}{\delta
 \tilde{\rho}^*_q(\mathbf{r}s,\mathbf{r'}s')}$
 where only the pairing energy in Eq.~(\ref{pairfunc})
contributes. 
The EDF considered here contains only local densities 
(see appendix~\ref{TDHFB_sphe}).
Therefore, the HF and anomalous fields are also local in space. 
Finally, it is  possible to recast the TDHFB equation~(\ref{eq:TDHFB_QPcomp}) as a set 
of Schroedinger like equations for the quasiparticle radial wave functions $u$ and~$v$
\begin{eqnarray}
  i \hbar \ddt
 \Svec{u_{nlj}}{v_{nlj}}  =
 \Smat{h_{lj}-\lambda}{\tilde{h}}{\tilde{h}^*}{-h_{lj}^*+\lambda}
 \Svec{u_{nlj}}{v_{nlj}}
\label{eq:TDHFB_qp}
 \end{eqnarray}
where $\lambda$ is the chemical potential (see section~\ref{sec:particle}).
The expressions of the fields $h(r)$ 
and~$\tilde{h}(r)$ and those of the various densities
entering the EDF solved in spherical symmetry
can be found in appendix~\ref{TDHFB_sphe}.

\subsection{Computational details}

The initial condition is obtained 
with the \textsc{hfbrad} code~\cite{hfbrad} which solves the static HFB 
equation in spherical symmetry. 
We have constructed a time-dependent version of this code 
to solve the TDHFB equation with the functional described 
in section \ref{sec:EDF}.
The set of equations~(\ref{eq:TDHFB_qp}) is solved iteratively
using a one-step predictor-corrector method~\cite{FKW}
and a truncation of the time propagator
\begin{eqnarray}
  U\left(t,t+\delta t\right)= \exp{\left( \frac{-i \delta t}{\hbar}\Hbogo
  (t+\delta t/{2})\right)}
\end{eqnarray}
 at $4$th order in $\delta t$.

Spatial derivatives are calculated in a discretized $r$-space using seven-points formula. 
The numerical accuracy of this method decreases increasing
quasi-particle energy. As a consequence, this approximate derivation
formula induces a small periodic variation of the HFB ground state
(of the order of few tens of keV) due to high energy quasiparticles.
We checked 
that this numerical artifact disappears if we
develop the wave-functions on a 
constant step Lagrange mesh~\cite{BayeLM}.
However, the latter method increases the numerical effort. 
The amplitude of these variations being linked 
to the mesh discretization, a mesh step of $0.15$~fm 
has been found to ensure a good numerical precision 
with a reasonable computational effort.
In particular, the energy is conserved up to $15$~keV
and deviations of the total number of particles are of the order of $10^{-7}$ 
in the present calculations. 
Though this latter value is one order of magnitude higher than in the TDHF case 
with the same numerical conditions, it is small enough to 
leave the observables of interest unaffected. 
Moreover, to avoid any unphysical contribution
due to the approximate spatial derivative 
in the evolution of observables, 
we subtract from their expectation values 
the one obtained without external field.
We checked that this procedure does not affect the physical content
of the spectra presented in Sec.~\ref{sec:results}.

Let us precise that we use hard box boundary conditions. 
The latter are not optimized for a proper treatment of the continuum
because they lead to a discretized quasiparticle spectrum.
In addition, particles which are reflected on the boundaries of the box 
may interact with the nucleus and induce unphysical effects
on the evolution of observables.
This problem has been tackled in TDHF with the help of 
absorbing boundary conditions~\cite{nak05, ABC_PRE}.
However, we did not found this technique to be appropriate 
to treat the TDHFB continuum because of the non-vanishing 
asymptotic nature of the upper component of the quasiparticle 
wave-functions in Eq.~(\ref{eq:qp_spher})~\cite{NPA422}. Though needed 
to refine description of unbound 
states~\cite{cQRPA}, further improvements of the boundary conditions 
are beyond the scope of this exploratory work.

We consider nuclei with magic proton numbers and 
excitations acting on neutrons only. 
Then the calculations include pairing for neutrons only.
The local pairing functional couples to high angular momenta and high energy quasiparticle states up to the energy cutoff of $80$~MeV.
In the calculations presented here, convergence 
of the static solutions have been obtained 
with a maximum total angular momentum for neutrons, 
for which the Bogoliubov transformation is achieved,
 of $j_c=19/2$ ($23/2$) for Oxygen (Calcium) isotopes respectively. 
 For Oxygen isotopes, a box radius $22.5$~fm was used, in 
order to be in conditions as close as possible to the discrete QRPA results of Ref.~\cite{Khan}. 
Calculations for Calcium isotopes have been performed in a box of radius $30$~fm.\\	
Both the mesh step, through the derivation formulae, and the maximum angular momentum, through 
the centrifugal part of the kinetic operator, constrain very much the time step for which the 
calculations are stable. The adopted time step used in these calculations are $0.003$ and 
$0.002$~fm/c for Oxygen and Calcium isotopes respectively. 

\section{Linear response framework for pairing excitations\label{sec:linear}}

\subsection{Linear response theory}

The linear response theory has been widely used with TDHF
to study collective vibrations in nuclei~\cite{blo79,str79,uma86,cho87,chi96,sim03,nak05,uma05,mar05,alm05,rei07,ste07}.
In this theory, one computes the time evolution of an observable 
\begin{equation}
\Delta Q(t)=\bra{\Psi (t)}\hat{Q}\ket{\Psi (t)} - \bra{0}\hat{Q}\ket{0}
\end{equation}
after an excitation induced by a small external potential
$\hat{V}_{ext}(t) = \epsilon \Fop \xi(t)$ 
on the ground state $\ket{0}$ of the system.
The parameter $\epsilon$ quantifies the intensity of the excitation.
It has to be small enough to ensure the linear regime, 
{ i.e.}, the amplitude $\Delta Q_{max}$ of the response 
must be proportional to $\epsilon$. In this study, the time dependence 
of the external potential is chosen to be a Dirac function $\xi(t)=\delta(t)$.
The excitation is then equivalent to a boost applied on the 
ground state at the initial time
\begin{equation}
\ket{\Psi(0)} = e^{-i\epsilon \Fop /\hbar} \ket{0}.
\label{eq:initial_condition}
\end{equation}

The response $\Delta Q(t)$ to this excitation can be decomposed
into various frequencies $\omega$ using
\begin{eqnarray}
R_{Q}(\omega) &=& \frac{-\hbar }{\pi \epsilon}\,
\int_{0}^{\infty} \!\!\! \Dt \,\,\, \Delta{Q}(t) \, \sin (\omega t).
\label{eq:spectral_response}
\end{eqnarray}
In the particular case where the operators used for the excitation 
and the observations are the same, i.e., 
 $\Fop = \hat{Q}$, Eq.~(\ref{eq:spectral_response}) gives the strength distribution
\begin{eqnarray}
R_{F}(\omega) &=& \sum_\alpha \, | \bra{\alpha} \Fop \ket{0} |^2 
\,\, \delta (\omega - \omega_\alpha)
\label{eq:strength}
\end{eqnarray}
where $ \bra{\alpha} \Fop \ket{0}$ is the transition amplitude between 
the ground state and the eigenstate $\ket{\alpha}$ of the Hamiltonian
 and $\hbar \omega_\alpha = E_\alpha-E_0$ is their energy difference.

When the excitation generates  a transition to neighboring nuclei, 
$E_0$ and $E_ \alpha$ are the energies of the 
ground and excited states of the final nucleus
if one keeps the static chemical potential in the Hamiltonian 
(see section 10.1 of Ref.~\cite{BlaizotRipka}).
For instance, in the case of an addition of two nucleons
on a nucleus of $A$ nucleons,
the energy of the mode reads 
$\hbar \omega_\alpha = E^{(A+2)}_\alpha-E_0^{(A+2)}$
where the ground state energy of the final
nucleus is approximated by $E_0^{(A+2)}=E_0^{(A)}+2\lambda$.
Note that this energy may differ from the one obtained
in a HFB calculation in the $A+2$ nucleus because of a possible 
rearrangement of the HFB field.
  
\subsection{Application to pairing vibrations}

Pairing vibrations of quantum numbers $0^+$ 
can be excited by two-nucleon transfer reactions 
and have been studied in the small amplitude limit 
within the QRPA framework~\cite{Khan}. 
A Hermitean pair-transfer operator is given by~\cite{BesBroglia}
\begin{eqnarray}
\hat{F}= \sum_{\nu}  \left(  f_\nu\,\crea_\nu \crea_{\bar{\nu}} 
+ f^*_\nu\, \anni_{\bar{\nu}} \anni_\nu \right)
\label{pairvibOBS}
\end{eqnarray}
where $\bar{\nu}$ denotes the time-reversed state of~$\nu$.

In this paper, we consider local excitations acting on neutrons only and, 
for the sake of simple notations, we do not write explicitly 
the isospin quantum number in the following.
The pair-transfer operator then writes in coordinate space
\begin{eqnarray}
\Fop = 
\int \!\!\! \mbox{d}\mathbf{r} \,\,\, f\ofR \left( \crea_{\mathbf{r},\downarrow} 
\crea_{\mathbf{r},\uparrow} 
+ \anni_{\mathbf{r},\uparrow} \anni_{\mathbf{r},\downarrow} \right).
\label{pairvibOBS2}
\end{eqnarray}
In this particular choice, the spatial distribution $f\ofR$ is real. 
Using Eq.~(\ref{eq:rhotilde}), the expectation value of $\hat{F}$ simply writes
\begin{equation}
\bra{\Psi (t)}\hat{F}\ket{\Psi (t)} = \frac{1}{2}\int\!\!\! \mbox{d}\mathbf{r} \,\,\,f(\mathbf{r})\,\, \left(\tilde{\rho}_0(\mathbf{r}; t)+\tilde{\rho}^*_0(\mathbf{r}; t)\right)
\label{eq:Fdet}
\end{equation}
where $\tilde{\rho}_0 = \sum_q\tilde{\rho}_q$.

To preserve spherical symmetry, we focus on monopole pairing modes,
requiring a radial dependence only, { i.e.},  $f\ofR\equiv f\ofr$.
We choose a Fermi-Dirac spatial 
distribution $f(r)=\left(1+\exp\left(\frac{r-R_c}{d}\right)\right)^{-1}$
where the parameters $R_c=(1.27~A^{1/3}+4)$~fm 
and $d=0.5$~fm are chosen to allow for pair transfer on the whole nucleus
on the one hand, and to remove unphysical high energy modes associated 
to pair creation outside of the nucleus on the other hand.
Finally, this excitation may change the number of neutrons
at the initial time. 
However, deviations are small
in the present calculations ($\sim 10^{-3}$ neutrons).

We choose to follow the excitation operator $\Fop$ itself 
to get its strength distribution defined in Eq.~(\ref{eq:strength}). 
We also decompose the excitation operator $\hat{F}= \sum_l \hat{F}_l$ 
into components of single particle angular momentum $l$
\begin{eqnarray}
\hat{F}_l &=& \sum_{nn'jm} \int \!\!\! \mbox{d} \mathbf{r} \,f(r)\,
 \langle nljm |\mathbf{r} \downarrow\rangle\,
\langle n'lj(-m) |\mathbf{r} \uparrow\rangle \nonumber \\
 && \times \crea_{nljm} \crea_{n'lj(-m)} + h.c.
 \label{eq:Fl}
\end{eqnarray}
where $h.c.$ denotes the {\it Hermitean conjugate} 
of the entire expression.
Computing the response of the $\hat{F}_l$
 helps us interpret the spectra in terms of specific 
quasiparticle excitations. 

We perform the calculations over a total time interval of $T=1200$~fm/c. 
In order to minimize the effects of the time gate on the Fourier transforms, 
we follow the protocol given in Ref.~\cite{ABC_PRE}, 
multiplying the observables by a time filter 
$\cos^2\left(\frac{\pi t}{2T}\right)$.
This procedure induces an additional width of~$\sim 1$~MeV.
 
\section{Results \label{sec:results}}

In this section, two-neutron transfer in several nuclei is studied.
To illustrate the method described in the previous section,
we first detail the analysis in $^{18}$O. 
Then, we present the results on neutron-rich Oxygen isotopes
and on $f$-shell Calcium isotopes.

\subsection{Detailed analysis on $^{18}$O}

\begin{figure}
\includegraphics[height=8cm,angle=-90]{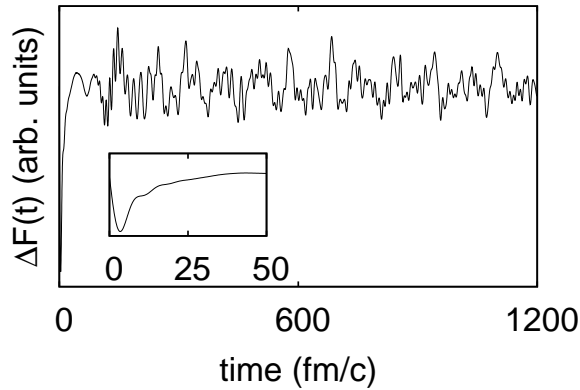}
\caption{Evolution of $\Delta F(t)$
after a pair transfer type
excitation on $\Ox{18}$. The inset shows the same quantity at early times.
\label{evolA_O18}}
\end{figure}

We apply the boost of Eq.~(\ref{eq:initial_condition}) 
in the linear regime on the HFB ground state of $\Ox{18}$.
The pair-transfer operator $\Fop$ is defined in Eq.~(\ref{pairvibOBS2}). 
The variation of the expectation value of $\Fop$, obtained from 
 Eq.~(\ref{eq:Fdet}), is plotted in 
{Fig.}~\ref{evolA_O18} as a function of time.
We observe a complex evolution due the excitation of several modes
at different energies. We see in the inset that a strong variation
of $\Delta F(t)$ occurs at early times 
because all modes are initially in phase~\cite{sim03}.

\begin{figure}
\includegraphics[height=8cm,angle=-90
]{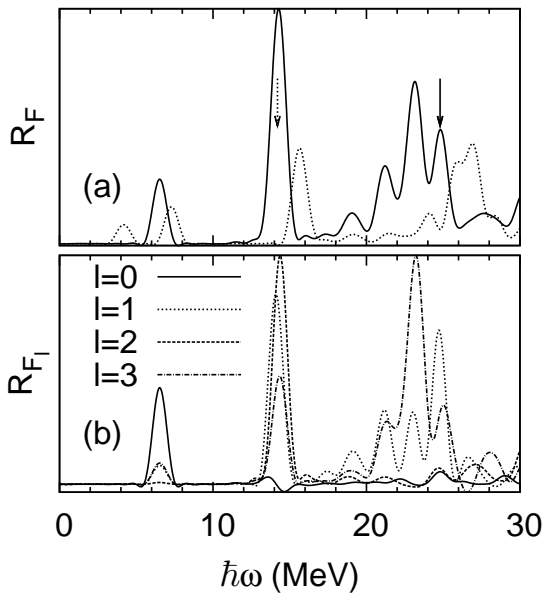}
\caption{Decomposition of the responses (in arbitrary units)
into frequencies $\omega$ for a two-neutron pair transfer 
excitation in $^{18}$O. (a) Strength distribution of $\hat{F}$ 
obtained from TDHFB (solid line) and the unperturbed approximation 
(dotted line).
The arrows indicate the $1p_{3/2}$ (solid) and $1p_{1/2}$ (dotted)
deep hole states. 
(b) TDHFB responses of $\hat{F}_l$ with $l=0, 1, 2$ and 3.
\label{TF_evolA_O18}}
\end{figure}

The evolution in {Fig.}~\ref{evolA_O18} is used to compute
 the strength distribution of $\hat{F}$ according to Eq.~(\ref{eq:spectral_response}) 
 and using the time filter procedure described in the previous section.
We have controlled that the extracted strength is independent of the 
excitation amplitude $\epsilon$.
 The resulting spectrum is shown in {Fig.}~\ref{TF_evolA_O18}(a) 
 in solid line. We see two separated peaks at 6.5 and 14.3 MeV 
 and several peaks between 20 and 26 MeV.

To understand the effect of the residual interaction 
which is included in TDHFB,
we have computed the so-called unperturbed response to the pair transfer excitation.
The latter is equivalent to Eq.~(\ref{eq:strength}) 
if one assumes that the states $\ket{\alpha}$ are two-quasiparticle 
excitations of the type $\ket{\mu \nu} =  \hbc_\nu \hbc_\mu\ket{0}$
where $\hbc_{\nu}$ creates a quasiparticle eigenstate 
of the static  HFB Hamiltonian on its ground state $|0\rangle$.
In this approximation, the energy of the transition
to the state $|{\mu \nu}\rangle$ 
is the sum of the quasiparticle energies 
$\hbar \omega_{\mu \nu} = e_\mu + e_\nu$.
To allow for a quantitative comparison between the strength distributions obtained
from TDHFB and unperturbed approximations, we compute the latter 
using the same time Fourier technique as for the TDHFB case. 
First, we determine the transition amplitudes $\langle{\mu \nu}|\hat{F}|0\rangle$
and then the time evolution of the observable $\hat{F}-\langle 0|\hat{F}|0\rangle$ 
within the unperturbed approximation. The latter reads
\begin{equation}
\Delta F^{0}(t) = \sum_{\mu \nu} |\langle{ \mu \nu}|\hat{F}|0\rangle|^2 
\sin (\omega_{{\mu\nu}} t).
\end{equation}
Finally, we apply exactly the same procedure 
 as for the TDHFB evolution to extract its strength distribution.
The resulting unperturbed spectrum is represented by a dotted line 
in {Fig.}~\ref{TF_evolA_O18}(a). We see clearly that the effect of 
the residual interaction is to increase the strength on the one hand
and, on the other hand, to shift down the positions of the peaks.

To get a deeper insight into the nature of the peaks, 
we decompose the response into components of the 
single particle orbital momentum $l$
using the observables $\hat{F}_l$ defined in Eq.~(\ref{eq:Fl}).
The response for $l=0,1,2$ and $3$ are plotted in 
{Fig.}~\ref{TF_evolA_O18}(b).
These spectra, together with the quasiparticle HFB spectrum,
allow us to characterize the peaks in terms of dominating 
two-quasiparticle excitations.
As one can see in {Fig.}~\ref{TF_evolA_O18}(b), the first peak located 
at $6.5$~MeV is associated to the $l=0$ component of $\hat{F}$.
It corresponds mainly to a pair transfer toward the almost empty $2s_{1/2}$ orbitals.

The next peak, located at $14.3$~MeV, is mainly a mixture of two contributions:
 the transfer of a pair towards $1d_{3/2}$ orbitals and
the removal of the $1p_{1/2}$ occupied neutrons
indicated by a dotted arrow in {Fig.}~\ref{TF_evolA_O18}(a). 
The fact that these two modes have the same energy is fortuitous.
(This is also the case at the unperturbed level.)
As we will see later, they are well separated in the other Oxygen isotopes.
We also see in Fig.~\ref{TF_evolA_O18}(b)
that there is a $l=3$ contribution to this peak
due to a coupling to $f_{7/2}$ orbitals in the continuum.

Let us now focus on the group of peaks at higher energies. 
As one can see in {Fig.}~\ref{TF_evolA_O18}(b), 
they are mostly populated by $l=1$ and 3 components.
In fact, the peaks between 20 and 24 MeV are mainly 
associated to the excitation of $f_{7/2}$ quasiparticle resonant states
while the peak at 24.7 MeV, indicated by a solid arrow in 
{Fig.}~\ref{TF_evolA_O18}(a), corresponds to the deep hole
$1p_{3/2}$ state.
Except for the latter contribution, which is due to the removal of 
two occupied neutrons, these peaks belong to the GPV~\cite{Khan}.
{Indeed, they correspond to excitations of resonant states
belonging to the next major shell and the enhancement of the 
strength as compared to the unperturbed spectrum is a sign of their 
collectivity}~\cite{BesBroglia}.

\subsection{neutron-rich Oxygen isotopes}

\begin{figure}
\includegraphics[height=8cm,angle=-90
]{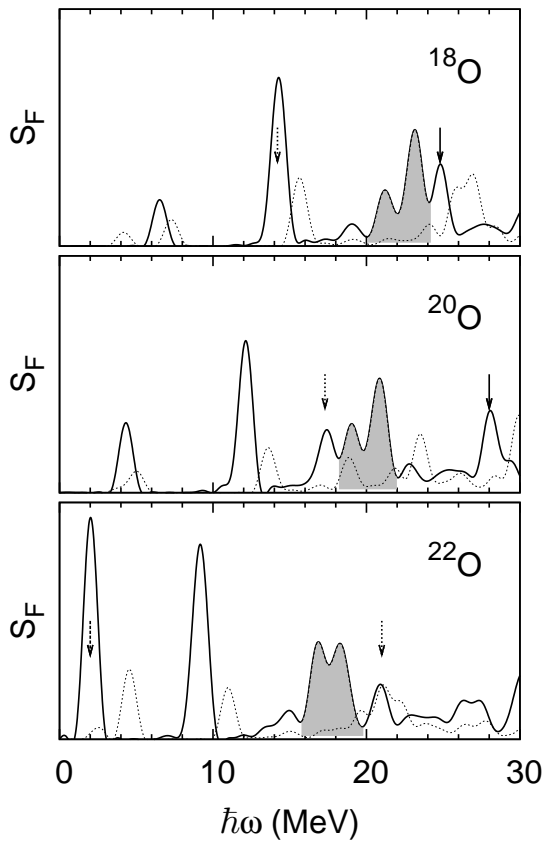}
\caption{Strength distributions of the two-neutron transfer operator $\Fop$ for 
$^{18,20,22}$O (in arbitrary units with the same scale on each plot).
TDHFB results (solid lines) and the unperturbed approximation (dotted lines) are shown.
The arrows indicate the $1p_{3/2}$ (solid) and $1p_{1/2}$ (dotted)
deep hole states and the $1d_{5/2}$ (dashed) pair removal. 
The filled regions correspond to GPV candidates (see text).
\label{GPV_Ox}}
\end{figure}

 \begin{table}
 \caption{\label{Ox_energy}
Energies and main quasiparticle contributions of the 
most important peaks appearing in the strength distribution of 
the two-neutron pair transfer operator $\hat{F}$ extracted from
TDHFB calculations for various Oxygen isotopes.
Numbers in parentheses are the centroids of the continuum-QRPA 
energies of Ref.~\cite{Khan}.
Labels in brackets indicate two-neutron removal contributions.}
 \begin{ruledtabular}
 \begin{tabular}{ccc}
nucleus & E (MeV) & main orbital contribution\\
\hline
$^{18}$O        & 6.5{\footnotesize\it (6.5)}     & $2s_{1/2}$ \\
                & 14.3{\footnotesize\it (14)}     & $1d_{3/2}$, $[1p_{1/2}]$ \\
                & 20-24{\footnotesize\it (21.5)}  & $f_{7/2}$ \\
                & 24.7                            & $[1p_{3/2}]$ \\
\hline
$^{20}$O        & 4.3{\footnotesize\it (4)}       & $2s_{1/2}$ \\
                & 12.1{\footnotesize\it (11)}     & $1d_{3/2}$ \\
                & 17.5                            & $[1p_{1/2}]$ \\
                & 18-22{\footnotesize\it (19)}    & $f_{7/2}$ \\
                & 28.0                            & $[1p_{3/2}]$ \\
\hline
$^{22}$O        & 2.0                             & $2s_{1/2}$, $[1d_{5/2}]$ \\
                & 9.2{\footnotesize\it (8)}       & $1d_{3/2}$ \\
                & 16-20{\footnotesize\it (16)}    & $f_{7/2}$ \\
                & 21.0                            & $[1p_{1/2}]$ \\
 \end{tabular}
 \end{ruledtabular}
 \end{table}

In addition to $^{18}$O, we also studied two-neutron 
pair transfer in $^{20,22}$O nuclei.
The spectra are shown in {Fig.}~\ref{GPV_Ox} 
while the energies and most important quasiparticle contributions 
to the main peaks are summarized in table~\ref{Ox_energy}. 
Comparing strength distributions obtained from TDHFB (solid lines) 
with the unperturbed approximation (dotted lines) 
in {Fig.}~\ref{GPV_Ox} leads to the same conclusions for all isotopes, 
{ i.e.}, an increase of the strength and a lowering of the peak energies
due to the TDHFB residual interaction. 
We also see in {Fig.}~\ref{GPV_Ox} that the energies 
of the $1p_{3/2}$ and $1p_{1/2}$ deep-hole states increase 
with the number of neutrons. The $1p_{3/2}$ peak in $^{22}$O
is located outside of the figure at 31.3 MeV.
The occupied single particle orbitals are indeed deeper
as compared to the Fermi level
when increasing the neutron number, corresponding to higher 
quasiparticle energies.
We also note that transitions associated to the $1d_{5/2}$ orbital appear only  
in the $^{22}$O spectrum. 
This is due to the fact that the operator $\hat{F}$ leaves 
the strongly paired levels almost unchanged~\cite{BesBroglia}, 
and then no significant strength is associated to 
addition and removal of nucleons into
a partially occupied level at the Fermi surface.
This is not the case in $^{22}$O where the 
 $1d_{5/2}$ single particle orbital is almost fully occupied.
 
Last but not least, we see that the GPV, indicated by a filled region, 
is present in the three isotopes with similar amplitudes.
In all cases, the most important contributions to the GPV 
are the excitation of $f_{7/2}$ quasiparticle resonant states.
In the present calculations, continuum states are discretized 
due to the finite size of the box, inducing a fragmentation of the GPV.
We also note that the GPV and the other two-neutron additional modes,
contrary to the removal ones, have a decreasing energy 
with the neutron number. This is due to the fact that the 
Fermi level is less deep for neutron-rich nuclei, 
which decreases the quasiparticle energy of states 
with small or zero occupation number.

Let us now compare the energies predicted by the present TDHFB calculations 
with the continuum-QRPA results of Khan {\it et al.}~\cite{Khan}, 
also reported in table~\ref{Ox_energy}. 
The latter have been computed for two-neutron additional modes only.
%
Compared to QRPA results, TDHFB calculations globally predict slightly higher energies 
when going to more neutron-rich nuclei. In the GPV region, the centroid energies 
are 0.5, 1 and 2 MeV higher with TDHFB for $^{18,20,22}$O respectively.
However, this overall agreement can be considered as good in regard to the 
differences between the two approaches.
%
Both calculations use the SLy4 Skyrme functional, but with different pairing schemes.
In the present work, we use a mixed volume-surface effective coupling.
In addition, our calculations are performed in wide quasiparticle energy 
and angular momentum windows, with cutoff values $E_c=80$ MeV 
and $j_c=19/2$ respectively, while the QRPA calculations 
have been performed in smaller windows
($E_{c}=50$~MeV and $j_c=9/2$) with a surface type pairing functional.
The parameters of the latter have been determined using a 
different prescription than ours~\cite{cQRPA}, in particular to reproduce 
the trend of the experimental gap in neutron-rich Oxygen isotopes.
Another possible source of discrepancies is the fact that 
the QRPA calculations of Ref.~\cite{Khan} do not take into account 
 the Coulomb and spin-orbit parts of the residual interaction whereas
the TDHFB approach uses the same EDF as the underlying HFB field.
This assumption may induce a slight shift in the energy of collective 
modes~\cite{colo04,per05,sil06}.

\subsection{Calcium isotopes}

\begin{figure}
\includegraphics[height=8cm,angle=-90
]{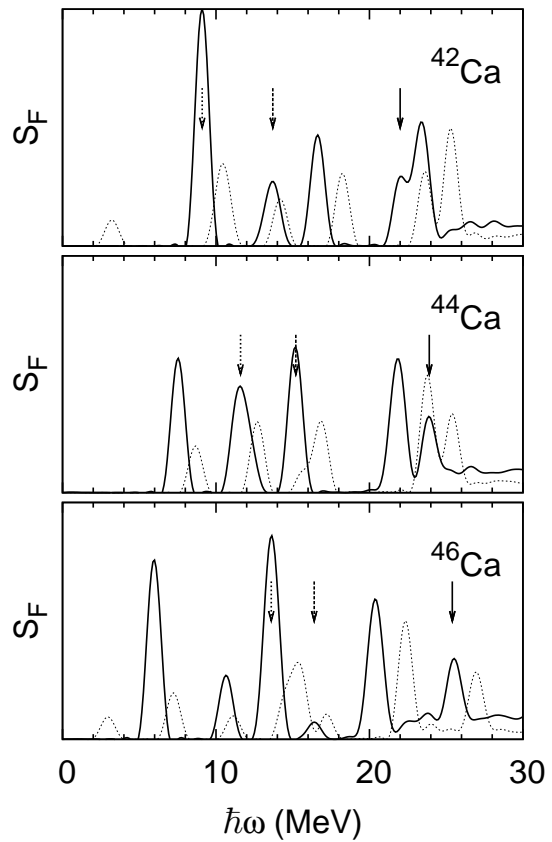}
\caption{Strength distributions of the two-neutron pair transfer operator $\Fop$ for 
 $^{42,44,46}$Ca (in arbitrary units with the same scale on each plot).
TDHFB results (solid lines) and the unperturbed approximation (dotted lines) are shown.
The arrows indicate the $1d_{5/2}$ (solid), $2s_{1/2}$ (dashed)
and $1d_{3/2}$ (dotted) two deep-hole states. 
\label{Ca0}}
\end{figure}

In this section we discuss two-neutron pair transfer on
$^{42,44,46}$Ca.
We have plotted in~{Fig}.~\ref{Ca0}
the corresponding TDHFB strength distributions (solid lines).
The spectra are roughly similar for the three isotopes.
They exhibit several discrete transitions to bound states
together with excitations of resonant two-quasiparticle states. 
The energy threshold for the latter can be estimated 
by twice the Fermi energy, {i.e.}, $2 E_{F}= 20.6$, 19.6 and 17.8 MeV
for $^{42}$Ca, $^{44}$Ca and $^{46}$Ca respectively.

We performed the same analysis as for the Oxygen isotopes, 
{ i.e.}, we decomposed the strength distribution in $l$-components 
which, together with the HFB quasiparticle spectra,
helped us assign the main quasiparticle contributions to each peak.
A summary of the results is given in table~\ref{Ca_energy} where
the transitions associated to the removal of two neutrons are indicated in brackets.

 \begin{table}
 \caption{\label{Ca_energy}
Energies and main quasiparticle contributions of the 
most important peaks appearing in the strength distribution of 
the two-neutron pair transfer operator $\hat{F}$ extracted from
TDHFB calculations for Calcium isotopes.
Labels in brackets indicate two-neutron removal contributions.\\
}
 \begin{ruledtabular}
 \begin{tabular}{ccc}
nucleus & E (MeV) & main orbital contribution\\
\hline
$^{42}$Ca 
& 9.1     & $2p_{3/2}$, $[1d_{3/2}]$ \\
& 13.7   & $2p_{1/2}$, $[2s_{1/2}]$ \\
& 16.6   & $1f_{5/2}$\\
& 22.0   & $[1d_{5/2}]$ \\
& 23.4   & $g_{9/2}$ \\
\hline
$^{44}$Ca 
& 7.5     & $2p_{3/2}$ \\
& 11.6   & $2p_{1/2}$, $[1d_{3/2}]$ \\
& 15.2   & $1f_{5/2}$, $[2s_{1/2}]$ \\
& 22.1   & $g_{9/2}$ \\
& 23.9   & $[1d_{5/2}]$ \\
\hline
$^{46}$Ca
& 6.0     & $2p_{3/2}$ \\
& 10.7   & $2p_{1/2}$ \\
& 13.6   & $1f_{5/2}$, $[1d_{3/2}]$ \\
& 16.4   & $[2s_{1/2}]$ \\
& 20.8   & $g_{9/2}$ \\
& 25.4   & $[1d_{5/2}]$ \\
 \end{tabular}
 \end{ruledtabular}
 \end{table}

For the three isotopes, the lowest mode is interpreted in terms of the addition of a 
neutron pair in the $2p_{3/2}$ orbitals.
In $^{42}$Ca, an additional $l=2$ quasiparticle component contributes to this mode
and corresponds to the removal of a $1d_{3/2}$ neutron pair. 
As for Oxygen isotopes, the appearance of removal modes (in brackets in 
table~\ref{Ca_energy}) together with additional modes at the same energy 
is fortuitous. In the Calcium isotopes, the removal modes are built of neutrons 
from the $s-d$ shell, the major shell below the Fermi energy.
As expected, one finally notes that energies of the removal 
(resp. additional) modes increase (decrease) with the neutron number. 
Although 2 $g_{9/2}$ quasiparticles excitations are forbidden below
the $2 E_F$ threshold in the unperturbed spectrum, it gets mixed
to the last bound 2 ($f_{5/2}$) quasiparticles excitation because of 
the residual interaction.

We have also plotted the unperturbed spectra 
in~{Fig}.~\ref{Ca0} (dotted lines). 
As in the case of Oxygen isotopes, 
we observe that the TDHFB residual interaction 
lowers the energies on the one hand and increases the 
strength on the other hand, though this second effect is 
less pronounced in the high energy part of the spectra. 
Here, the pairing residual interaction
is not strong enough to gather high energy states together 
and to generate a well identified collective pairing vibration
in the continuum.
In a pure independent particles shell model picture,
the last occupied neutron level of these Calcium isotopes is 
the $1f_{7/2}$ orbital.
Then, one expects the GPV to be built mainly on low-lying $g_{9/2}$
resonant quasiparticle states. However, as we clearly
see in~{Fig}.~\ref{Ca0}, the strength associated
to the two-quasiparticle excitation of $g_{9/2}$ levels,
located at $23.4$, $22.1$ and $20.8$~MeV for 
$\Ca{42}$, $\Ca{44}$ and $\Ca{46}$ respectively, 
is only slightly enhanced by the TDHFB residual interaction.
We checked that employing other parameterizations 
of the spatial distribution $f(r)$ of the excitation operator 
in Eq.~(\ref{pairvibOBS2}) does not alter these conclusions.
We note that this lack of collectivity of the $g_{9/2}$ has already
been observed in hole pairing giant resonances studies 
within a more schematic formalism~\cite{her85}.

\section{Conclusion \label{sec:conclusion}}

We solved the TDHFB equation
in coordinate space with spherical symmetry
for the evolution of a single nucleus in an external field.
For the normal part of the energy density functional, 
we used the SLy4 Skyrme functional.
For its pairing part, we chose a local density dependent functional.
Special care has been taken regarding the convergence of the static HFB solutions
and the energy and particle number conservations in the TDHFB calculations.

As a first application, we studied $0^+$ pairing modes excited 
by a two-neutron pair transfer type operator.
The linear response theory has been used to compute the strength 
distributions of this operator in $^{18,20,22}$O and $^{42,44,46}$Ca nuclei.
Both transitions to bound states and to the continuum are observed in all nuclei.
In particular, the GPV is observed in all Oxygen isotopes
whereas no significant enhancement of the strength 
due to dynamical pairing correlations appears in the continuum 
of the studied Calcium isotopes. In the latter, the $g_{9/2}$ 
quasiparticle excitations are not collective enough to generate a GPV.

A detailed comparison with previous QRPA calculations have been performed 
in the Oxygen isotopes. Though there is a good agreement 
for the most stable isotope, we find slightly higher energies for the
pairing vibrations when going to more neutron-rich nuclei.
Different pairing schemes and implementations of the residual interaction 
in both calculations are invoked to explain these differences.

In addition, there is room for a better treatment of the continuum, 
for instance in the spirit of continuum-QRPA calculations,
but such an improvement is not straightforward. Indeed, one cannot 
extrapolate the absorbing boundary conditions used in TDHF calculations
to the TDHFB case because of the delocalized upper components 
of the Bogoliubov spinors.

Finally, TDHFB calculations are much more demanding 
in terms of computational time than standard TDHF calculations, 
by about two orders of magnitude in the one dimensional case. 
However, thanks to recent increase of computational power, 
some of the standard TDHF applications to nuclear structure and reactions
should be repeated with the inclusion of dynamical pairing correlations
in the framework of TDHFB.

\begin{acknowledgments} 
We thank K. Bennaceur for his help on the \textsc{hfbrad} code 
and the associated formalism. 
Discussions with M. Bender, T. Duguet, H. Flocard, E. Khan, 
D. Lacroix and V. Rotival are gratefully acknowledged. 
One of us (B.A.) also thanks R. Broglia and G. Ripka for useful discussions 
at the early stage of this work. 
Calculations have been performed at the Centre de Calcul CC-IN2P3.
\end{acknowledgments}

\appendix
\section{Densities and fields in spherical symmetry \label{TDHFB_sphe}}


Each energy term entering Eq.~(\ref{TDEDF}) 
can be written as a spatial integral of an energy density, i.e.,
$  \mathcal{E} = \int \!\!  \mbox{d}\mathbf{r} \,\, \mathcal{H}( \mathbf{r})$,
 which depends only on local densities and currents.
For a spherically symmetric system, the densities entering the SLy4 
and local density dependent pairing functionals are
the radial part of the local particle $\rho_q(\mathbf{r})$, 
anomalous $\tilde{\rho}_q(\mathbf{r})$, 
kinetic  $\tau_q(\mathbf{r})$, matter current $\jmat_q(\mathbf{r})$ 
and spin-orbit current $\Jso_q(\mathbf{r})$ (both 
oriented along the radial unit vector $\mathbf{e}_r)$ densities of isospin $q$.
Introducing the notation $\alpha \equiv \{n,l,j,q\}$,
these densities write
\begin{eqnarray}
  \rho_q(\mathbf{r}) &=&
  \sum_{nlj} 
  K_j(r)
  |{v}_{\alpha}(r)|^2 
  , \nonumber \\
  \tilde{\rho}_q(\mathbf{r})
  &=& - \sum_{nlj} 
  K_j(r)
 {v}_{\alpha}^*(r) u_{\alpha}(r) 
  , \nonumber \\
  \tau_q(\mathbf{r}) &=& \sum_{nlj} 
  K_j(r)
  \left[ \left| 
  \left(\frac{\partial}{\partial_r}  - \frac{1}{r}\right){v}_{\alpha}(r) \right|^2 \right. 
  \nonumber \\
  &&+ \left. \frac{l(l+1)}{r^2} |{v}_{\alpha}(r)|^2 \right] 
  , \nonumber \\
  \jmat_q(\mathbf{r}) &=& 
  \sum_{nlj}
  K_j(r)
 \,  \mbox{Im}\left[{v}_{\alpha}(r)\left(\frac{\partial}{\partial_r} 
  - \frac{1}{r}\right){v}^*_{\alpha}(r)\right] \mathbf{e}_r
  , \nonumber \\ 
  \Jso_q(\mathbf{r}) &=& 
  \sum_{nlj} 
  \frac{K_j(r)}{r}
   \left[j(j+1)-l(l+1)-\frac{3}{4}\right]|{v}_{\alpha}(r)|^2 \mathbf{e}_r
  , \nonumber
\end{eqnarray}
where  $K_j(r)=\frac{ 2 j + 1 }{ 4 \pi r^2 }$.
We also define isoscalar densities as the sum
of proton and neutron densities, e.g., 
$\rho_0(\mathbf{r})= \rho_p(\mathbf{r})+\rho_n(\mathbf{r})$.
Omitting the dependence in $\mathbf{r}$  to simplify the notations,
the different parts of the functional can be written
\begin{eqnarray}
  \mathcal{H}_{kin} &=& 
  \frac{\hbar^2}{2m} \left( 1 - \frac{1}{A} \right) \tau_0
  , \nonumber \\
  \mathcal{H}_{Sk} &=& 
  \sum_{k=n,p,0} \left\{ \, \left(C^{\rho}_k+C_k^{\rho^\alpha}\rho_0^\alpha\right) \rho_k^2 
  + \, C_k^{\Delta \rho} \rho_k \Delta \rho_k \right.  \nonumber \\
  &\,& \left. + C_k^{\tau} \left( \rho_k \tau_k - \jmat_k^2\right) + C^{\Jsob} \rho_k \nabla.\Jso_k \,\, \right\} 
  , \nonumber \\
  \mathcal{H}_{pair} &=& 
  \frac{g}{4}\left(1-\left(\frac{\rho_0}{\rho_c}\right)^\gamma\right) \,
   \sum_{q=p,n} \tilde{\rho}_q^* \, \tilde{\rho}_q 
  , \nonumber \\
  \mathcal{H}_{Coul} &=& 
  V^{dir}_c \rho_p - \frac{3}{4} e^2\left(\frac{3}{\pi}\right)^{1/3} \rho_p^{4/3} 
  , \nonumber
\end{eqnarray}
where $  V^{dir}_c\left(r\right) = \frac{e^2}{2}
  \int \mbox{d}\mathbf{r}' \frac{\rho_p(r')}{|\mathbf{r}-\mathbf{r}'|}$ is the direct
  Coulomb field.
The factor (1-1/A) in the kinetic part is the so-called one-body 
center of mass correction. 
The $\jmat^2$ term ensures Galilean invariance~\cite{eng75}. 

The coefficients $C$ are related to the usual
 Skyrme coefficients by (see, e.g.,~\cite{Cranking})
\begin{eqnarray*}
  C^{\rho}_0            &=& \frac{t_0}{2} \left(1+\frac{x_0}{2}\right)  , \\ 
  C^{\rho}_{n,p}        &=& -\frac{t_0}{2} \left(x_0+\frac{1}{2}\right) , \\ 
  C^{\rho^\alpha}_0     &=& \frac{t_3}{12}\left(1+\frac{x_3}{2}\right)  , \\ 
  C^{\rho^\alpha}_{n,p} &=& -\frac{t_3}{12}\left(x_3+\frac{1}{2}\right) , \\ 
  C^{\tau}_0     &=& \frac{t_1}{4} \left(1+\frac{x_1}{2}\right) 	
	    +\frac{t_2}{4} \left(1+\frac{x_2}{2}\right)                 , \\ 
  C^{\tau}_{n,p} &=& -\frac{t_1}{4} \left(x_1+\frac{1}{2}\right) 	
	    +\frac{t_2}{4} \left(x_2+\frac{1}{2}\right)                 , \\ 
  C^{\Delta \rho}_0     &=& -\frac{3t_1}{16} \left(1+\frac{x_1}{2}\right) 
            +\frac{ t_2}{16} \left(1+\frac{x_2}{2}\right)               , \\ 
  C^{\Delta \rho}_{n,p} &=& \frac{3t_1}{16} \left(x_1+\frac{1}{2}\right) 	
            +\frac{ t_2}{16} \left(x_2+\frac{1}{2}\right)               , \\ 
  C^{\Jsob}   &=&  -\frac{W_0}{2}.
\end{eqnarray*}

The fields entering Eq.~(\ref{eq:TDHFB_qp}) write
\begin{eqnarray}
h_{ljq}\left(r\right) &=& 
  - \frac{\partial}{\partial r }  M_q\left(r\right) \frac{\partial}{\partial r }
  + I_{q}(r) \frac{\partial}{\partial r }
  + V_{ljq}\left(r\right)  \nonumber \\
\tilde{h}_q\left(r\right) &=& 
   \frac{g}{2}\left[1-\left(\frac{\rho_0(r)}{\rho_c}\right)^\gamma\right]\tilde{\rho}_q(r) \nonumber
\end{eqnarray}
with
\begin{eqnarray}
M_q &=& 
    \frac{\hbar^2}{2 m} 
    + C^{\tau}_0 \rho_0 + C^{\tau}_q \rho_q 
    \nonumber \\
I_q &=& 
    2 i \left( C^{\tau}_0 \, \jmat_0 + C^{\tau}_q \, \jmat_q \right) \cdot \mathbf{e}_r
    \nonumber \\
V_{ljq} &=& U_q +
    \frac{l(l+1)}{r^2} M_q  
    + \frac{1}{r}\left(\frac{\partial}{\partial r } M_q\right)   - \frac{I_q}{r}
    \nonumber \\
&~& -
    \frac{j(j+1)-l(l+1)-3/4}{r} \,C^{\Jsob} \left(\frac{\partial}{\partial r }\left(\rho_0+ \rho_q \right)\right)
   .
    \nonumber \\
U_q &=& \sum_{k=q,0} 
    \left[ 
      2~\left(C_k^{\rho} + C^{\rho^\alpha}_k \rho_0^\alpha \right)\rho_k  \right.
    \nonumber \\
&~&
    \left. 
        \left.      + \alpha C^{\rho^\alpha}_k  \rho_0^{\alpha-1} (\rho_p^{2}+ \rho_n^{2})
+        C^{\tau}_k \tau_k  
    \right.     \right. 
    \nonumber \\
&~& + 
 \left.         2~C^{\Delta \rho}_k  \Delta \rho_k 
        + C^{\Jsob} \nabla \cdot \Jso_k
+        i  C^{\tau}_k  \, \nabla \cdot \jmat_k \right]
    \nonumber \\
&~& -
    \frac{g}{4} \gamma \frac{\rho_0^{\gamma - 1}}{\rho_c^\gamma} \left(|\tilde{\rho}_p|^2+|  \tilde{\rho}_n|^2\right).      \nonumber 
%
%
\end{eqnarray}



\end{document}